\documentclass[runningheads]{llncs}
\usepackage{graphicx}
\usepackage{amsmath}
\usepackage{amssymb}
\usepackage{multirow}
\usepackage{subfig}
\usepackage{makeidx}
\makeindex

\begin{document}
\title{Cascaded Framework for Automatic Evaluation of Myocardial Infarction from Delayed-Enhancement Cardiac MRI}
\titlerunning{Cascaded Framework for Myocardial Pathology Segmentation}
%
\author{Jun Ma}\index{Ma, Jun}
\authorrunning{J. Ma}
\institute{Department of Mathematics, Nanjing University of Science and Technology \email{junma@njust.edu.cn}}
%
%
\maketitle              
\begin{abstract}
Automatic evaluation of myocardium and pathology plays an important role in quantitative analysis of patients suffering from myocardial infarction. In this paper, we present a cascaded convolutional neural network framework for myocardial infarction segmentation and classification in delayed-enhancement cardiac MRI.
Specifically, we first use a 2D U-Net to segment the whole heart, including the left ventricle and the myocardium. Then, we crop the whole heart as a region of interest (ROI). Finally,  a new 2D U-Net is used to segment the infraction and no-reflow areas in the whole heart ROI. The segmentation method can be applied to the classification task where the segmentation results with the infraction or no-reflow areas are classified as pathological cases.
Our method took second place in the MICCAI 2020 EMIDEC segmentation task with Dice scores of 86.28\%, 62.24\%, and 77.76\% for myocardium, infraction, and no-reflow areas, respectively, and first place in classification task with an accuracy of 92\%.

\keywords{Segmentation \and Myocardial Pathology \and Cascaded Framework.}
\end{abstract}

\section{Introduction}
Quantitative assessment of myocardial viability is essential in the diagnosis and treatment management for patients suffering from myocardial infarction (MI). Cardiac magnetic resonance (CMR) is particularly used to provide imaging anatomical and functional information of the heart, such as the delayed-enhancement (LGE) CMR sequence which visualizes MI.

\begin{figure}
\center
\includegraphics[scale=0.35]{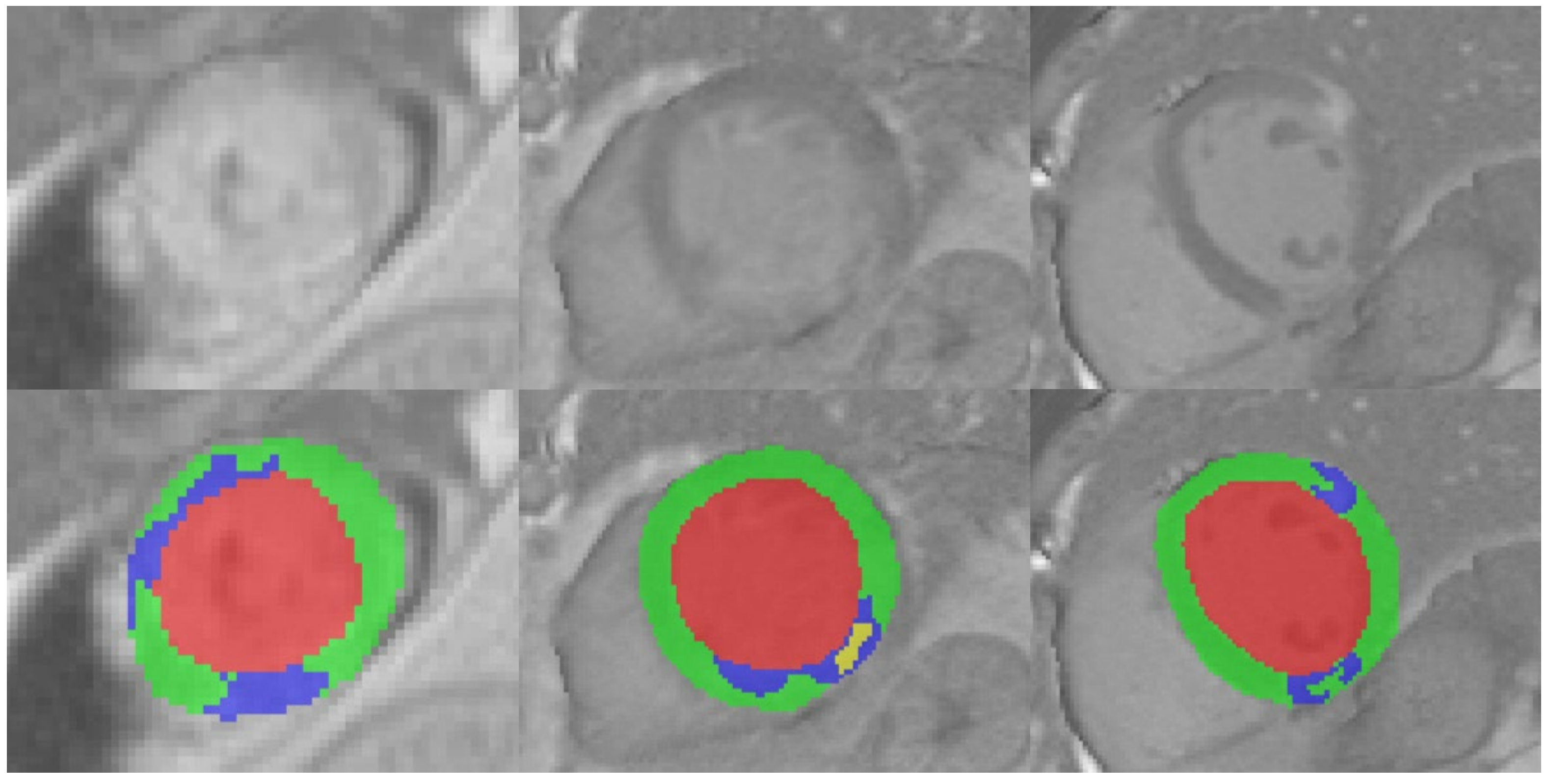}
\caption{Visual examples of different myocardial infraction delayed-enhancement cardiac MR images. The 1st row and the 2nd row are the original image and ground truth, respectively.  In the 2nd row, the red, green, blue,and yellow color denote left ventricle,  healthy myocardium, infraction, and no-reflow, respectively.} \label{fig:ImgExample}
\end{figure}

One of the important tasks is to segment the myocardium into different regions, including normal myocardium, infarction, and no-reflow from multi-sequence CMR dataset.
Manual annotation is generally time-consuming, tedious and subjects to inter- and intra-observer variations. Thus, a fully automatic segmentation method is highly desired in clinical practice. Figure~\ref{fig:ImgExample} presents some images from different myocardial infraction cases and the corresponding left ventricle, healthy myocardium, infraction, and no-reflow annotations. It can be observed that the intensity appearances vary significantly among different cases, and both infraction and no-reflow areas have ambiguous boundaries and low contrast.
Thus, it is very challenging to automatically segment them.

To the best of our knowledge, most CMR segmentation related studies focus on the left ventricle/atrium, right ventricle, and myocardium segmentation (\cite{ACDC2018,ZhuangWHO2019,chen2020CardiacReview,ma2020LGAC}), little work has been done in the fully automatic cardiac pathology segmentation (\cite{ZhuangPAMI18CMRSeg,LiLei2020MIA,lilei2020jointSeg,Zhai2020MyoPS,ma2020MyoPS}). To advance the development of myocardial infraction image analysis, a joint segmentation and classification challenge, automatic evaluation of myocardial infarction from delayed-enhancement cardiac MRI (EMIDEC, \url{http://emidec.com/}), was organized in MICCAI 2020.

The two main objectives of the EMIDEC challenge are first to classify normal and pathological cases from the clinical information with or without DE-MRI, and secondly to automatically detect the different relevant areas (the myocardial contours, the infarcted area and the permanent microvascular obstruction area (no-reflow area)) from a series of short-axis DE-MRI covering the left ventricle. The segmentation allows us to make a quantification of the MI, in absolute value (mm3) or percentage of the myocardium.
The paper presents our method details for the EMIDEC challenge.

\section{Method}
\begin{figure}
\center
\includegraphics[scale=0.34]{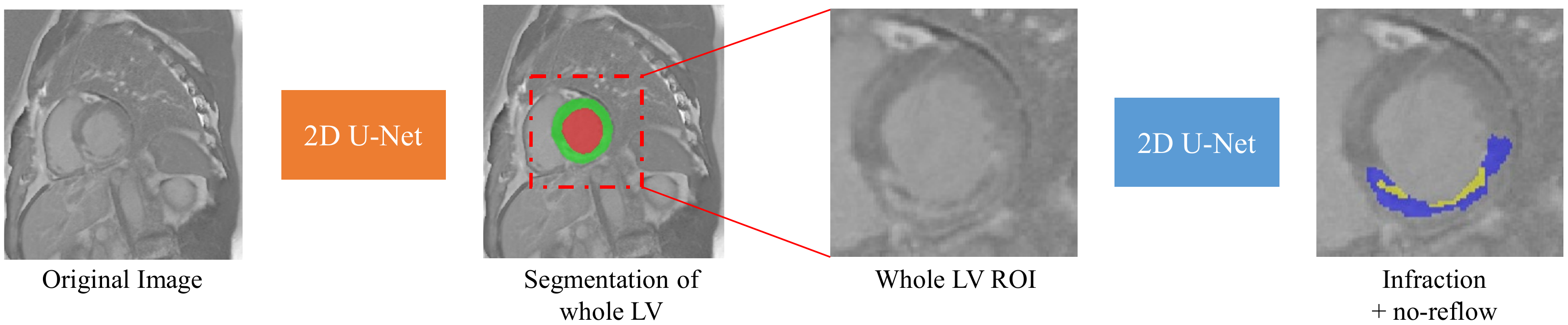}
\caption{Pipeline of the proposed method. we first use a 2D U-Net to segment the whole LV (left ventricle), including LV blood pool and myocardium. Then, we crop the LV region of interest (ROI). Finally, a new 2D U-Net is used to segment the infraction and no-reflow areas.} \label{fig:pipeline}
\end{figure}

This paper focuses on both healthy and pathology pathologic myocardium segmentation from the delayed-enhancement cardiac MRI.
One of the main challenges is how to exploit rich and reliable information regarding the pathological as well as morphological information of the myocardium.
To this end, we design a cascaded framework that comprises two 2D U-Net to segment the left ventricle and myocardium, and the pathology regions, respectively. Figure~\ref{fig:pipeline} presents the whole pipeline of the proposed method. Specifically, the proposed method contains three steps\footnote{In step 1 and step 3, the networks are trained end-to-end, while the whole framework is not end-to-end.}:
\begin{itemize}
    \item Step 1 (whole LV segmentation). Train a 2D U-Net~\cite{UNet2d} on the original CMR images to segment the whole LV (including left ventricular blood pool and myocardium);
    \item Step 2 (creating ROI). Crop LV region of interest (ROI) from the original CMR images based on the segmentation results in step 1. In this way, the unrelated background can be excluded;
    \item Step 3 (infraction and no-reflow segmentation). Train a new 2D U-Net to segment the infraction and no-reflow from the ROI images.
\end{itemize}

\section{Dataset and Training protocols}
The EMIDEC challenge dataset provides 100 cases for the training and 50 cases for the testing~\cite{lalande2020EMIDEC-Data}. For the classification tasks, additional 50 testing cases are provided and participants are required to submit the classification results within one hour.
In particular, Every training and test case represents a DE-MRI exam of the left ventricle. An exam (i.e. a case) consists of a series of 5 to 10 short-axis slices covering the left ventricle from the base to the apex. The ground-truths (contours of the relevant areas) will be provided with the training dataset. The training set with full ground-truth will comprise 100 cases (67 pathological cases, 33 normal cases) randomly selected among the 150 subjects. The testing set is made of data from 50 subjects (33 pathological cases, 17 normal cases), all different from those in the training set.

During preprocessing, we apply z-score to normalize the image intensity, and resample all the images to the same spacing $10.0\times1.458\times1.458$ $mm^3$. For the image, we apply three-order spline interpolation for in-plane voxels and nearest neighbor interpolation for out-of-plane voxels. For the ground truth, we convert the label to one-hot encoding and apply linear interpolation for in-plane voxels and nearest neighbor interpolation for out-of-plane voxels.
We employ nnU-Net \cite{nnunet20} as the main network without any modification in architecture. Since the CMR data has a large slice thickness, 2D U-Net is more suitable for this task. During the training of the first U-Net, the patch size is $256\times224$ and the batch size is 16. During the training of the second U-Net, the patch size is $66\times66$, and the batch size is 32.
The loss function is the sum between Dice loss \cite{dice2016} and cross entropy. We use stochastic gradient descent with momentum to optimize the networks.
Each model is trained on a TITAN V100 GPU.
During testing, we use a five-fold ensemble to predict each testing case.

\section{Results and Discussion}
\subsection{Cross-validation segmentation results}
Table~\ref{tab:CV-Heart} presents the quantitative cross-validation results of the first U-Net.
The U-Net can achieve very high accuracy for the whole LV, which can ensure the cropped ROI can cover most of the LV and also the lesions.

\begin{table}[!h]
\caption{Quantitative segmentation results of the left ventricle, myocardium and whole LV on the training set.}\label{tab:CV-Heart}
\centering
\begin{tabular}{cccc}
\hline
Metrics      & Left Ventricle (LV) & Myocardium (Myo) & Whole LV (LV+Myo) \\ \hline
Dice (\%) & 93.47 $\pm$ 2.06        & 85.38 $\pm$ 3.94      & 95.51 $\pm$ 1.83       \\ \hline
\end{tabular}
\end{table}

\begin{figure}
\center
\includegraphics[scale=0.5]{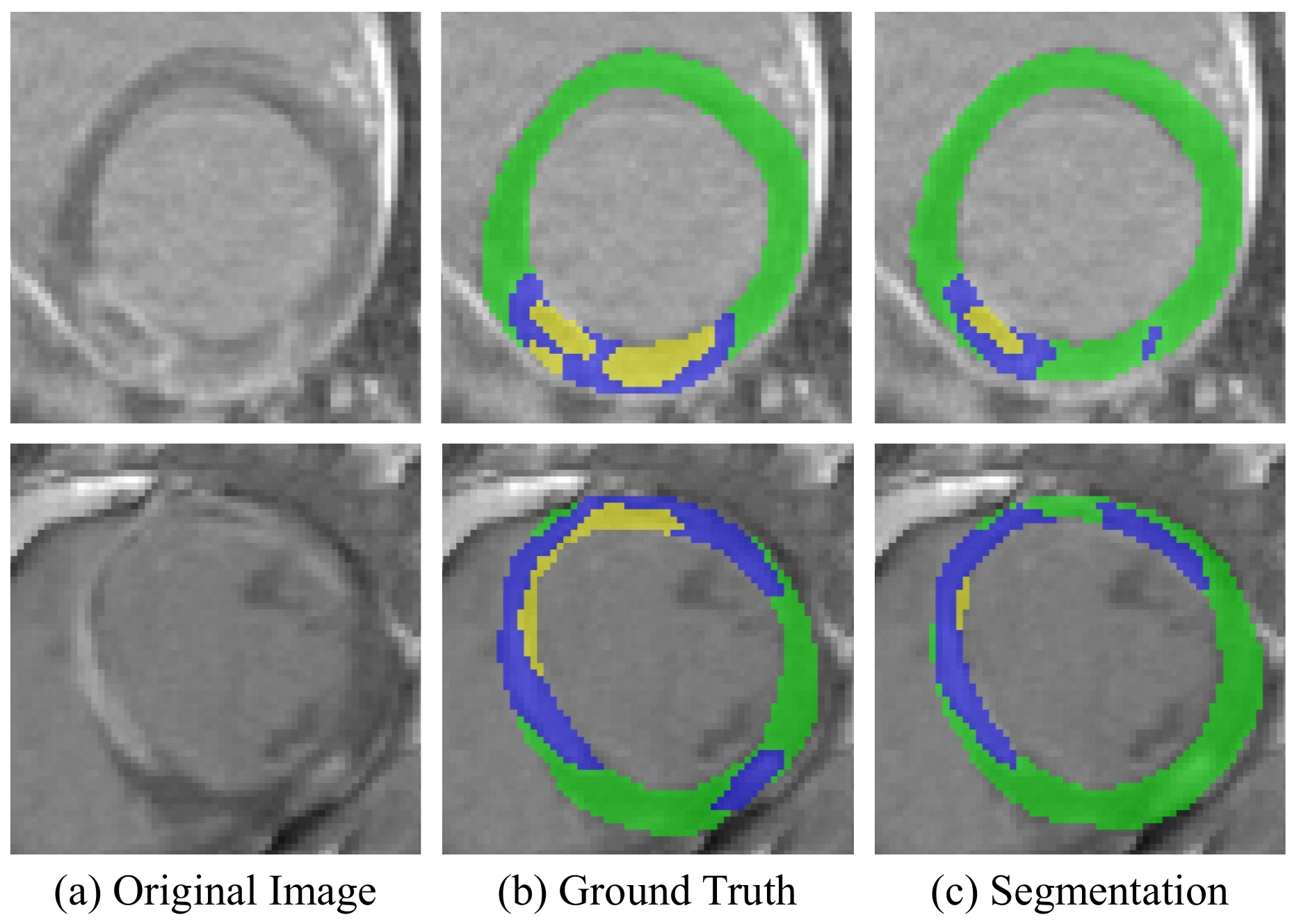}
\caption{Visualized segmentation results of the myocardium (green), infraction (blue) and no-reflow areas (yellow).} \label{fig:seg}
\end{figure}

Table~\ref{tab:CV-lesion} shows the quantitative segmentation results of the infection and the no-reflow. The sensitivity of all lesions is significant worse than the corresponding specificity, indicating that most segmentation results are right but many lesions are missed by the proposed method.
Figure~\ref{fig:seg} presents the visualized segmentation results. We can find that most of the missed infraction and no-reflow areas have low contrast and weak boundaries, which are very challenging to segment.

\begin{table}[!h]
\caption{Quantitative segmentation results of the infraction and no-reflow on the training set.}\label{tab:CV-lesion}
\centering
\begin{tabular}{lccc}
\hline
\multicolumn{1}{c}{Metrics} & Infraction  & No-reflow   & Infraction + No-reflow \\ \hline
Dice (\%)                & 57.96 $\pm$ 16.64 & 77.19 $\pm$ 31.33 & 59.23 $\pm$ 17.85            \\
Sensitivity (\%)         & 53.61 $\pm$ 19.23 & 74.02 $\pm$ 34.88 & 54.16 $\pm$ 20.20            \\
Specificity (\%)         & 99.21 $\pm$ 0.62  & 99.94 $\pm$ 0.10  & 99.26 $\pm$ 0.61             \\ \hline
\end{tabular}
\end{table}

\subsection{Testing set segmentation results}
We evaluate the proposed method on the official testing set with 50 cases. Table~\ref{tab:test} presents the quantitative results. Our results ranked the second place on the overall segmentation ranking leaderboard (\url{http://emidec.com/leaderboard}).

\begin{table}[!htbp]
\caption{Quantitative segmentation results of the testing set. NR stands for the no-reflow.}\label{tab:test}
\centering
\begin{tabular}{llcc}
\hline
Targets                     & Metrics          & Results & Subranks     \\
\hline
\multirow{3}{*}{Myocardium} & DSC (\%)              & 0.8628 & 2  \\
                            & Volume Difference ($mm^3$)        & 10153 & 2   \\
                            & Hausdorff Distance ($mm$)               & 14.31 & 3   \\
\hline
\multirow{3}{*}{Infarction} & DSC (\%)              & 0.6224 & 3 \\
                            & Volume Difference ($mm^3$)        & 4874  & 4  \\
                            & Volume Difference Ratio (\%) & 3.50 & 3 \\
\hline
\multirow{3}{*}{Re-flow}    & DSC (\%)              & 0.7776 & 3  \\
                            & Volume Difference ($mm^3$)       & 829.7 & 2  \\
                            & Volume Difference Ratio (\%) & 0.49 & 2 \\
\hline
\end{tabular}
\end{table}

\subsection{Testing Set Classification}
The final segmentation results can be used for classifying the cases in normal or pathological. In particular, if the segmentation of one case does not have lesions (infraction or no-reflow) or the number of the lesion voxels is less than 10, it will be classified as a normal case. Otherwise, it will be classified as a pathological case.

The classification was organized as an on-site challenge where the participants had one hour to run their methods on their own laptop to classify 50 addition testing cases between normal and pathologic cases.
Table~\ref{tab:Class} presents the quantitative classification results.
Our results ranked the first place on the classification ranking leaderboard with an accuracy of 92\%.

\begin{table}
\centering
\caption{Classification Leaderboard.}
\label{tab:Class}
\begin{tabular}{lcc}
\hline
Teams            & Accuracy & Rank  \\
\hline
Ma               & 92\%     & 1     \\
Lourenco et al.  & 82\%     & 2     \\
Ivantsits et al. & 78\%     & 3     \\
Sharma et al.    & 62\%     & 4     \\
\hline
\end{tabular}
\end{table}

\section{Conclusion}
This paper presents a simple fully automatic method for myocardium and pathology segmentation and classification from enhanced cardiac MR images. Experiments on the MICCAI 2020 EMIDEC challenge dataset show that the proposed method can achieve promising results, which ranked the 2nd place in segmentation task and the 1st place in classification task.
In future, we would improve the learning ability of the network to be more sensitive to the lesions.

\section*{Acknowledgment}
The authors of this paper declare that the segmentation method they implemented for participation in the EMIDEC challenge has not used any pre-trained models nor additional MRI datasets other than those provided by the organizers.
The author also highly appreciates the organizers of automatic Evaluation of Myocardial Infarction from Delayed-Enhancement Cardiac MRI (EMDEC 2020) for their public dataset and organizing the great challenge.

%
%
\bibliographystyle{splncs04}
\bibliography{REF}

\end{document}